\documentclass[aps,prl,notitlepage,%
longbibliography,twocolumn]{revtex4-1}
\usepackage{natbib}
\usepackage{graphicx}
\usepackage{amssymb}
\usepackage{amsmath}
\usepackage{ifthen}
\usepackage{braket}
\usepackage{xcolor}
\usepackage{bm}
\usepackage{bbm}
\usepackage{hyperref}
\usepackage{comment}
\usepackage{siunitx}
\usepackage{float}
\usepackage{dcolumn}
\newcolumntype{d}[1]{D{.}{.}{#1}}
\usepackage{subfigure}
\usepackage[greek,english]{babel}

\newcommand{\dd}{\mathrm{d}}
\newcommand{\ii}{\mathrm{i}}
\newcommand{\ee}{\mathrm{e}}

\newcommand{\barn}{{\overline{n}}}
\newcommand{\barz}{{\overline{z}}}

\newcommand{\hate}{\hat{\mathrm{e}}}

\definecolor{garrosgreen}{rgb}{0.1, 0.4, 0.1}
\definecolor{dartmouthgreen}{rgb}{0.05, 0.5, 0.06}
\definecolor{duelferred}{rgb}{0.7, 0.2, 0.1}
\definecolor{cambridgeblue}{rgb}{0.1, 0.3, 1.0}
\definecolor{oxfordblue}{rgb}{0.05, 0.2, 0.7}

\begin{document}

\title{Reply to Comment on ``Revisiting the divergent multipole expansion of
atom-surface interactions: Hydrogen and positronium, 
\greektext{}a\latintext{}-quartz,
and physisorption'' \\{}
[arXiv:2501.14803 [physics.atom-ph]]}

\author{Ulrich D. Jentschura}
\affiliation{Department of Physics and LAMOR, Missouri University of Science and
Technology, Rolla, Missouri 65409, USA}

\begin{abstract}
We present a Reply to the Comment 
by G. L. Klimchitskaya, arXiv:2501.14803 [physics.atom-ph].
It is shown that the criticism formulated in the
Comment fails to appreciate
recently obtained results for the upper limit
of the short-range expansion of atom-surface interactions,
and that the application of our results
to physisorption is based on a valid extension
of Lifshitz theory to the physisorption range,
which can be accomplished by 
refining the concept of the atom-surface distance 
with the help of a reference-plane that takes
the response function of the solid into account.
Some details on the calculation of the 
reference-plane are recalled from the literature.
\end{abstract}

\maketitle

{\em General Remarks.---}In 
the Comment~\cite{Kl2025comment}, the author claims
that expressions for the multipole corrections to 
atom-surface interactions are indicated incorrectly,
both in the short- and long-range asymptotic regimes,
in contradiction with those dictated by the Lifshitz theory.

These claims are misleading because of the following
reasons.  {\em (i)} Our article is 
based on an {\em extension} of Lifshitz theory, 
which extends the applicability of the $1/z^3$ short-range
expressions to very close approach of atom and surface,
upon a suitable modification developed by Zaremba and 
Kohn~\cite{ZaKo1976} as discussed in our
article~\cite{Je2024multipole} and ramified below.
{\em (ii)} The ranges of validity of the short-range,
and long-range expressions indicated in Ref.~\cite{Je2024multipole}
are indicated correctly for zero-temperature (non-thermal)
field theory, which is an underlying assumption
of the paper. Modifications for finite temperature 
and for particular atomic species
have recently been discussed by us in Ref.~\cite{DaUlJe2024}.

In order to understand the misunderstanding
on which the Comment~\cite{Kl2025comment} is based,
we must distinguish between the assumptions of the 
original Lifshitz theory and the modifications 
derived by Zaremba and Kohn~\cite{ZaKo1976} for close approach,
but first, let us concentrate on Lifshitz 
theory~\cite{Li1955,LaLi1960vol8,DzLiPi1961spu}.

{\em Considerations within Lifshitz theory.---}Within 
Lifshitz theory, as correctly stated
in the Comment, the solid ``wall'' can be assumed to 
fill the half-space $z \leq 0$, and the wall
material is treated as a continuous medium.
The atom is placed at the $z$ coordinate.
No attempt is made within Lifshitz theory, to 
analyze, say, the effect of individual atomic layers 
near the surface of the material.
Hence, the condition for the short-range cutoff of the 
applicability of Lifshitz-theory expressions,
namely, $z \gg a_0$ where $a_0$ is the Bohr radius,
is correctly indicated in Ref.~\cite{Je2024multipole},
in the text following Eq.~(22) of Ref.~\cite{Je2024multipole}.

Starting from a range $z \gg a_0$, the $1/z^3$ nonretarded 
interaction potentials are applicable~\cite{Li1955,LaLi1960vol8,DzLiPi1961spu,%
DzLiPi1961advphys,Pa1974,SpTi1993,TiSp1993a,TiSp1993b,Je2024multipole}.
The Comment~\cite{Kl2025comment} then claims that the 
short-range regime in which Lifshitz theory is 
applicable for
\begin{equation}
\label{differ}
d \ll z \ll \lambda_0 \,,
\end{equation}
where $d$ is an interatomic distance in the wall material,
and $\lambda_0$ is a characteristic absorption
wavelength of the wall material.
For single crystals with a simple atomic structure,
which are the subject of our study~\cite{Je2024multipole},
$d$ is of the order of the
Bohr radius, and hence the above statements
for the lower limit of the applicability of the
short-range regime are confirmed.
The discussion arises with regard to the upper limit 
of the applicability of Lifshitz theory, where the above
Eq.~\eqref{differ} differs from the condition
$z \ll a_0/\alpha$ indicated 
in the text following Eq.~(22) of Ref.~\cite{Je2024multipole}.
(Here, $\alpha$ is the fine-structure constant.)

First, it is interesting to observe that, 
in Eq.~(2) of Ref.~\cite{CrGuRe2019},
the applicability of the short-range expansion
is assumed to be 
\begin{equation}
\label{correct}
a_0 \sim d \ll z \ll \frac{a_0}{\alpha} \sim \lambda_A \,,
\end{equation}
where $\lambda_A$ is the wavelength of the first 
atomic (dipole) transition, in accordance with 
our paper.
We have recently shown~\cite{DaUlJe2024} that the 
condition for the upper limit of the 
short-range regime is more precisely given by 
\begin{equation}
\label{right}
z \ll \chi \, \frac{a_0}{\alpha} \,,
\qquad 
\chi = \sqrt{ \frac{ \alpha_{\rm a.u.}(0) }{Z } } \,,
\end{equation}
where $\alpha_{\rm a.u.}(0)$ is the static polarizability
of the atom (in atomic units)
and $Z$ is the number of electrons
of the atom. Up to the prefactor $\chi$, this
result agrees with the condition $z \ll a_0/\alpha$ 
indicated in Eq.~\eqref{correct} 
above and in the text following Eq.~(22) of Ref.~\cite{Je2024multipole}.
The parameter $\chi$ is of order unity 
for ground-state hydrogen and positronium but 
can differ from unity for other atomic species, 
as detailed in Fig.~6 of Ref.~\cite{DaUlJe2024}, for other elements.
We have performed extensive numerical calculations
to support the condition~\eqref{right}, 
as detailed in Ref.~\cite{DaUlJe2024}.

The condition $z \ll \lambda_0$ mentioned in the Comment~\cite{Kl2025comment}
otherwise indicates an excessively large upper limit for the 
range of applicability of the short-range
expressions of Lifshitz theory. 
Let us consider an extreme case, in principle favorable
for an extended range of the short-range expansion,
namely, metastable helium
(in the $2{}^3S_1$ state),
interacting with a gold surface.
Based on Eq.~\eqref{differ},
estimates for the transition region to the
retarded regime have
5een indicated by the author of the Comment~\cite{Kl2025comment}
to be in the range of
$z \leq 150 \, {\rm nm} \approx 3000 \, a_0 \approx \lambda_0$
in the text following Eq.~(3)
in Sec.~II of the author's work in Ref.~\cite{CaKlMoZa2005},
and in Sec.~16.3.4 and Sec.~16.4.2 of the author's work
in Ref.~\cite{BoKlMoMo2009}.
However, even for this extreme case, as shown in Ref.~\cite{DaUlJe2024},
the short-range expansion breaks down earlier, 
namely, at $z \approx 1300 a_0$ according to 
the result indicated in the text following
Eq.~(16c) of Ref.~\cite{DaUlJe2024}.
For other atomic species like hydrogen,
the short-range expansion even breaks down
at $z \approx 201 a_0$,
as also indicated in the text following
Eq.~(16c) of Ref.~\cite{DaUlJe2024}.
We can conclude that the estimate~\eqref{differ} 
indicates an upper limit for the applicability of the 
nonretarded regime which is excessively large.
As shown in Ref.~\cite{DaUlJe2024}, retardation
effects set in earlier than implied by the condition~\eqref{differ}.

The Comment then indicates that the long-range, $1/z^4$ 
asymptotics are valid in the regime
\begin{equation}
\lambda_0 \ll z \ll \frac{\hbar c}{k_B T} \approx 7.6 \, \mu{\rm m} \,.
\end{equation}
Upon the modification $\lambda_0 \to \chi \, a_0/\alpha$, we
agree with this statement, {\rm provided one uses 
thermal field theory for the derivation of the 
atom-surface interaction}.
As recalled in Appendix A of Ref.~\cite{DaUlJe2024},
for finite temperature, 
there exists a very-long range nonretarded
tail proportional to $1/z^3$, which is due to effects described
by thermal field theory (contributions of the first Matsubara
frequency). The effect vanishes at zero temperature,
i.e., in the limit $T \to 0$, which implies
\begin{equation}
\chi \frac{a_0}{\alpha} \sim \frac{a_0}{\alpha} \ll z \ll \infty \,,
\end{equation}
as indicated for the applicability 
of the long-range expansion
in the text following Eq.~(22) of Ref.~\cite{Je2024multipole}.
As evident from all initial formulas delineated
in Ref.~\cite{Je2024multipole}, our work~\cite{Je2024multipole}
is based on non-thermal field theory ($T \to 0$).

{\em Considerations beyond Lifshitz theory.---}
It may not be universally known in the 
atomic-physics community that considerable efforts 
have been invested, over the past decades, 
into an extension of Lifshitz theory to 
close approach of atom and surface, even 
down to the range of physisorption
(i.e., into the range
of $\sim$ 4--7 a.u., see Eq.~(2.39) of Ref.~\cite{ZaKo1976}).
The calculation of the reference-plane position $z_0$,
which allows one to extend Lifshitz theory 
into the range $z \sim d$, can meanwhile be regarded
as textbook material (see Chap.~2 of Ref.~\cite{BrCoZa1997}
and Chap.~6 of Ref.~\cite{Li1997}).
The general paradigm is as follows:
In order to improve the description of 
atom-surface interactions, one determines a reference-plane 
which can loosely be associated with the centroid
of the induced charge distribution in the solid
and replaces the factor $1/z^3$ from Lifshitz
theory by the expression $1/(z-z_0)^3$ (this idea
has been expanded on in Ref.~\cite{ZaKo1976}).
Pioneering considerations in regard to the
calculation of the reference-plane have been reported
in the early 1970s (see Fig.~3 of Ref.~\cite{LaKo1971}
and Sec.~IV and Fig.~2 of Ref.~\cite{LaKo1973}).
The theory has been significantly advanced in the 
seminal paper~\cite{ZaKo1976}.
Refinements of the theory have been 
discussed in Refs.~\cite{Fe1976,BaBaRa1979,Fe1981,PeZa1984}.
Sum rules fulfilled by the dynamic 
reference-plane positions have been discussed
in Ref.~\cite{PeAp1983,PeZa1985}.
The reference-plane position $z_0$ for 
interactions of helium atoms with 
metals has been calculated in Ref.~\cite{Li1986},
and checked against the sum-rule approach~\cite{PeZa1985}.
A more recent example for an application
of the theory can be found in Ref.~\cite{TaRa2014}.
The extension of Lifshitz theory into the 
range $z \sim d$ can thus be regarded as a very-well-established
concept, and we shall expand on this point in the 
following, including a treatment of the repulsive 
tail due to nearest-neighbor interactions,
which can be described density-functional calculations (see 
Refs.~\cite{Gr2004,Gr2006,Gr2010,SiSt2008,%
ChASJo2012,SiAmGrAn2012}).

In order to ramify this statement, let us 
briefly consider the general arguments outlined 
in Ref.~\cite{ZaKo1976}.
We start from Eq.~(2.15) of Ref.~\cite{ZaKo1976},
which we recall for convenience
(in the following derivation, atomic units are employed,
with $\hbar = |e| = 1$, $c = 1/\alpha$, $\epsilon_0 = 1/(4 \pi)$,
see Chap.~ 2 of Ref.~\cite{JeAd2022book}, and the atom-surface
interaction energy is denoted as $E$):
\begin{equation}
E = -\frac{2}{\pi} \, \int_0^\infty \dd u \,
\alpha(\ii u) \, F(\ii u, Z) \,.
\end{equation}
Here, $\alpha(\ii u)$ is the atomic 
(dipole) polarizability at imaginary angular frequency 
$\ii u$. The $F$ function is given in Eq.~(2.16) of Ref.~\cite{ZaKo1976},
\begin{align}
F(\ii u, Z) =& \; \frac{2 \pi^2}{L^2} \,
\sum_{\vec q} \exp(- 2 q Z) \,
\nonumber\\
& \; \times
\int \dd z \, \int \dd z' \, \ee^{q z} \, \ee^{q z'} \,
\chi_s(z, z', \vec q, \vec q, \ii u)
\nonumber\\
=& \; \frac{2 \pi^2}{L^2} \,
\sum_{\vec q} \exp(- 2 q Z) \, f(q, \ii u) \,,
\end{align}
where $\chi_s(z, z', \vec q, \vec q', \ii u)$ is the (nonlocal) response
function of the solid (evaluated at $\vec q' = \vec q$), as defined in Eqs.~(2.9) 
and~(2.16) of Ref.~\cite{ZaKo1976}, and $f(q, \ii u)$ 
is defined in the obvious way.
The response function  $\chi_s(z, z', \vec q, \vec q, \ii u)$
is Fourier transformed with respect to the 
$xy$ coordinates, so that $\vec q = q_x \hate_x + q_y \hate_y$ is 
a momentum vector in the $xy$ plane.
Now, we use the following relation,
derived in Eq.~(2.27) and (2.34) of Ref.~\cite{ZaKo1976},
\begin{align}
f(q, \ii u) =& \;
\frac{q}{2 \pi} \,
\int \dd z \, \delta \barn_0(z, \ii u) \,
\left[ 1 + 2 q \, \barz(\ii u) \right]
\nonumber\\
=& \;
\frac{q}{2 \pi} \,
\frac{\epsilon(\ii u) - 1}{\epsilon(\ii u) + 1} \,
\left( 1 + 2 q \barz(\ii u) \right) \,,
\end{align}
where $\barn_0(z, \ii u)$ is the dynamic (frequency-dependent)
averaged (over a lattice spacing) induced charge density 
in the solid [see Eqs.~(2.23) and~(2.25) of Ref.~\cite{ZaKo1976}], 
and $\barz(\ii u)$ is associated with,
but not necessarily equal to, the centroid of the 
induced charge density in the solid [see 
the comprehensive discussion
in the text following Eq.~(2.28) of Ref.~\cite{ZaKo1976}].
One decisive identification not explicitly 
mentioned in Ref.~\cite{ZaKo1976} concerns the 
continuum limit over the allowed momentum modes
in the (large) surface area $L^2$ of the solid,
\begin{equation}
\sum_{\vec q} = \frac{L^2}{(2 \pi)^2} \,,
\end{equation}
a relation whose three-dimensional generalization is extremely 
useful in counting vacuum modes in laser physics 
[see Eq.~(1.1.39) of Ref.~\cite{ScZu1997} and Eq.~(3.16) of 
Ref.~\cite{JeKe2004aop}]. It then takes only a few more 
steps to show that
\begin{align}
E =& \; -\frac{2}{\pi} \,
\int_0^\infty \dd u \,
\alpha(\ii u) \, 
\biggl\{ \frac{2 \pi^2}{L^2} \,
\sum_{\vec q} \exp(- 2 q z) \, \biggr\}
\nonumber\\
& \; \times
\biggl\{
\frac{q}{2 \pi} \,
\frac{\epsilon(\ii u) - 1}{\epsilon(\ii u) + 1} \,
\left( 1 + 2 q \barz(\ii u) \right)
\biggr\}
\nonumber\\
=& \; -\frac{1}{\pi} \,
\int_0^\infty \dd u \,
\alpha(\ii u) \,
\int \dd^2 q \, \frac{q}{2 \pi} \,
\exp(- 2 q z) 
\nonumber\\
& \; \times
\frac{\epsilon(\ii u) - 1}{\epsilon(\ii u) + 1} \,
\left( 1 + 2 q \barz(\ii u) \right)
\nonumber\\
=& \;
-\int_0^\infty \dd u \, \frac{\alpha(\ii u)}{\pi (2 z)^3} \,
\frac{\epsilon(\ii u) - 1}{\epsilon(\ii u) + 1} \,
\left( \Gamma(3)  + \frac{ \Gamma(4) }{z} \barz(\ii u) \right)
\nonumber\\
= & \; - \frac{1}{4 \pi z^3}
\int_0^\infty \dd u \, \alpha(\ii u) \,
\frac{\epsilon(\ii u) - 1}{\epsilon(\ii u) + 1} \,
\left( 1 + \frac{ 3 }{z} \barz(\ii u) \right)
\nonumber\\
= & \; - \frac{C}{z^3}
\left( 1 + \frac{ 3 z_0 }{z} \right)
= - \frac{C}{(z - z_0)^3} + \dots \,,
\end{align}
where $\Gamma$ denotes the Gamma function,
\begin{equation}
C = \frac{1}{4 \pi} \int_0^\infty \dd u \, \alpha(\ii u) \,
\frac{\epsilon(\ii u) - 1}{\epsilon(\ii u) + 1} 
\end{equation}
is the $C_3$ coefficient from Lifshitz theory and 
\begin{equation}
z_0 = \frac{1}{4 \pi C} \int_0^\infty \dd u \, \alpha(\ii u) \,
\frac{\epsilon(\ii u) - 1}{\epsilon(\ii u) + 1} \,  \barz(\ii u) 
\end{equation}
is the reference-plane position.
(One notes that in Eq.~(35) of Ref.~\cite{JeMo2023blog},
the notation $C_{30} = C_3$ is used, in view of 
the existence of logarithmic terms in atom-surface
interactions, which implies the necessity of an additional index.)
Having calculated $z_0$, one can extend the Lifshitz theory 
for close approach according to the 
replacement $1/z^3 \to 1/(z - z_0)^3$.
For the example systems studied in 
Ref.~\cite{TaRa2014}, and with reference to Eq.~\eqref{correct},
we observe that $z_0$ is of the same 
order-of-magnitude as the Bohr radius $a_0$
(numerical results are given in Table~III of Ref.~\cite{TaRa2014}).
The same is true for the numerical results given in Table~I
of Ref.~\cite{Li1986}.

The established, accepted procedure within the 
community, manifest in Ref.~\cite{TaRa2014}
(see also Refs.~\cite{ZaKo1976,PeBuEr1996,Gr2004,Gr2006,Gr2010,SiSt2008,%
ChASJo2012,SiAmGrAn2012})
expands on this idea by separating the problem in the physisorption
range into two parts: one part, which is calculated 
on the basis of density-functional theory 
and a second part,
which takes care of the interactions 
with all other atoms in the bulk.
The nearest-neighbor part 
(in the sense of the ``repulsive nearest-neighbor repulsion
energy'') is addressed on the basis 
of density-functional theory.
The interaction with all other atoms in the sample
takes place at a distance commensurate with the 
assumptions of Ref.~\cite{ZaKo1976} and 
can thus be taken into account in terms
of a reference-plane-corrected Lifshitz theory.
It is for the latter part that 
the quadrupole correction can become numerically 
significant, as shown in Ref.~\cite{TaRa2014}.
As detailed in the derivation surrounding Eqs.~(64) and~(65)
of our Ref.~\cite{Je2024multipole},
by using our results, communicated in Ref.~\cite{Je2024multipole},
within the formalism developed in Ref.~\cite{TaRa2014},
one obtains improved agreement 
for Kr on Cu(111) and Ar on Pd(111)
upon the inclusion of the quadrupole correction.
E.g., for Kr on Cu(111), a quadrupole correction of 15\,meV
is added to a leading dipole term of 86\,meV 
and a contact part of 20\,meV, the latter being calculated
based on density-functional theory, to yield a final
result of 121\,meV for the physisorption energy.
This calculation
employs the formalism of Ref.~\cite{TaRa2014}
(see the entries in Table~III of a Ref.~\cite{TaRa2014})
and illustrates the relevance of the quadrupole 
correction.  The quadrupole correction, for Kr on Cu(111),
amounts to $15/(86 + 20) = 14.2\%$ and is thus 
phenomenologically relevant.

We have calculated, in Eqs.~(62) and
(63)~of our paper~\cite{Je2024multipole},
the quadrupole corrections to the 
atom-surface interaction of positronium 
and hydrogen with $\alpha$-quartz at $z = 10 \, {\rm a.u.}$, 
and shown that the quadrupole effect is phenomenologically 
relevant in this distance range, even if 
$z = 10 \, {\rm a.u.}$ is larger than 
the distance range relevant to physisorption.
The $z$ coordinate in Eqs.~(62) and
(63)~of our paper~\cite{Je2024multipole}
should be interpreted in terms of the distance
of the $z$-coordinate of the hydrogen and positronium atoms
with respect to the reference plane $z_0$,
whose calculation has not been considered in Ref.~\cite{Je2024multipole}
for the respective systems at hand.

{\em Conclusions.---}In summary, we disagree with the 
conclusions of Ref.~\cite{Kl2025comment}:
The distance ranges for the short-range and long-range
approximations of the atom-wall interaction 
indicated in Ref.~\cite{Je2024multipole}
(see also Ref.~\cite{CrGuRe2019})
have been verified against recent work (Ref.~\cite{DaUlJe2024}),
and the extensions of Lifshitz theory 
necessary for very close approach (physisorption range)
constitute textbook material
(see Chap.~2 of Ref.~\cite{BrCoZa1997}
and Chap.~6 of Ref.~\cite{Li1997}).
Applications to physisorption confirm the importance
of the quadrupole correction.

{\em Acknowledgments.---}The author acknowledges
helpful conversations with Professor~Carsten A.~Ullrich.
This research was supported by the National
Science Foundation (Grant PHY--2110294).

\end{document}